\documentclass[prl,aps]{revtex4}
\usepackage{graphicx}

\begin{document}

\title[ Reduction of One Loop Feynman Diagrams in Scalar Field Theory]{Reduction of One Loop Feynman Diagrams in Scalar Field Theory}
 { Vestnik Leingradskogo Gosudarstvennogo Universiteta, 2, N 16, page 17, 1979}

\author{A.G. Izergin and  V.\ E.\ Korepin}
\affiliation{Lenigrad department of Steklov mathematical institute}

\begin{abstract}
This is a historical note.  In 1979 we wrote a paper in Russian Journal , see \cite{scan}.
We  considered massive scalar quantum filed theory. One loop Feynman diagrams were evaluated. Theorem was proved that one loop diagram with many internal lines [more then dimension of space-time] can be expressed in terms of one loop diagram with number of internal lines equal to the dimension of space-time [multiplied by tree diagrams].
This is translation in English.
\end{abstract}


\maketitle

\section{ Introduction}
The paper is devoted to derivation of reduction formula for one-loop diagrams in scalar quantum field theory.
Our formula reduce calculation of arbitrary one-loop diagram to calculation of one-loop diagram with number of internal lines equal to dimension of space time [{\bf which we denote by n}].
In case n=2, the formula is  known, see \cite{kt,ak}.  It helps to establish the structure of S-matrix in two dimensional models of quantum field theory, see  \cite{ak,zam}.
 The reduction formula for higher n  \cite{mel,pet}  is less known. Here we rigorously prove the reduction formula for arbitrary dimension of space-time.  The method is a generalization a standard procedure, which helps  to evaluate  the Feynman integral with respect to time component of momentum $k_0$: analytical continuation of integrand into complex plane of
 $k_0$ and subsequent calculation by   residues.  It turns out that in one-loop diagram one can use integration by  residues sequentially with respect to all components of momentum $k$ for any n.
  We hope that this method can be generalized to other models of quantum field theory and multi-loop diagrams.
 
 In section 1 we  formulate the reduction rule (formula  \ref{rr}) for one-loop diagrams for arbitrary $n$. In section 2 we prove the reduction formula. Section 3 contains technical details.  
 
 \section{1. Reduction Rule}
 Let us consider Euclidean quantum field theory of scalar fields with polynomial interaction in $n$ dimensional space-time.
 The number of fields and their masses are arbitrary positive numbers $m_i >0$. We shall denote the number of internal lines in one-loop diagram by $N \ge n$.
 For calculation of these diagrams it is enough to consider only triple vertexes.  The figure of one-loop diagram with triple vertexes can be found on the second page of the original \cite{scan}, it looks like a circle. An external momentum coming into vertex $i$ is denoted by $p_i$.
 Let us denote by $p_{ki}$  the total external momentum coming into the diagram from vertex number $i$ to  $k+1$. The direction from vertex $i$ to vertex $k$ is counted in clockwise direction.
 $$ p_{ki}=p_{i+1}+p_{i+2}+\ldots +p_k =-p_{ik}$$ of cause $p_{ii}=0$. We shall denote the momentum in the internal line going from vertex $i$ to vertex $i+1$ by $k_i$. So we have $k_j=k_i+p_{ji}$.
 Feynman integral corresponding to the diagram is
 \begin{equation} \label{Fi}
 \int d^nk_i \Pi_{j=1}^N\left( k_j^2+m_j^2 \right)^{-1}= D^{(n)}_N \left(  m_1^2, \ldots m_N^2 ; p_1, \ldots p_N\right)
 \end{equation}
 Since $N\ge n$ the integral is convergent. It does not depend on which of $k_i$ we chosen as integration variable. Since all $m_j^2 >0$ the integral $D^{(n)_N}$ is a smooth, decaying function of all external momenta and masses.   We assume that external momenta are in generic position: 
  \begin{itemize}
 \item  for $n=N$ any subset of $n-1$ momenta from the set of all $\{ p_i\}$ are linear independent
 \item For $N>n$ we consider  $n+1$ partial sums: $p_{a_2 a_1}, p_{a_3 a_2}, 
 \ldots p_{a_{n+1} a_n}, p_{a_1 a_{n+1}} $ (here $a_i>a_j$ if $i>j$) such that 
 $$p_{a_2 a_1}+ p_{a_3 a_2}+ p_{a_{n+1} a_n}+ p_{a_1 a_{n+1}}=0 $$
 Any $n$ of these vectors are linear independent.
   \end{itemize}
{\bf  We shall call these conditions {\it A}.}
  
 Now let us formulate the {\it reduction rule} for the diagram $D^{(n)}_N$  introduced in formula \ref{Fi}.
The rule express   $D^{(n)}_N$ as  a sum of 2 ${N\choose n}$ terms. Each term is a product of one-loop diagram with $n$ internal lines  $D^{(n)}_n$ by a tree diagram: 
\begin{eqnarray}   \label{rr}
&D^{(n)}_N = \sum _{a_1<a_2<\ldots <a_n}  \left[ \frac{1}{2}\int  d^n k_{a_1} \prod_{i=1}^n  \left( k_{a_i}^2+m_{a_i}^2 \right)^{-1} \right] \times  \\ \nonumber
& \times \{ \prod_{l\neq a_i} \left( (k_l^+)^2+m_l^2 \right)^{-1} + \prod_{l\neq a_i} \left( (k_l^-)^2+m_l^2 \right)^{-1}  \}  
\end{eqnarray}\
 Here $k_{a_i}= k_{a_1} +p_{a_i a_1} $ also  $k^{\pm}_{l}= k^{\pm}_{a_1} +p_{l a_1} $ and the values $k^{\pm}_{a_1}  $ are defined as points of joint pole of propagators $  \left( k_{a_i}^2+m_{a_i}^2 \right)^{-1}$.
 We can also define them as solutions of system of $n$ equations:
 \begin{equation} \label{jp}
 k_{a_i}^2+m_{a_i}^2=  (k_{a_1} +p_{a_i a_1} )^2 +m_{a_i}^2    =0  \qquad  \mbox{here} \qquad  i=1,2,\ldots , n
 \end{equation}
 The system has two solutions, see formula  \ref{sol} . \par
 {\it  The formula \ref{rr} is a multi-dimensional analogue of the standard Cauchy's  formula for evaluation of an integral of a  holomorphic function by residuals. 
 One can say that in formula  \ref{rr}  we moved $(N-n)$ propagators $(k_l^2 + m_l^2)^{-1}$ out of the integral at the point of joint pole  of $n$ propagators  $(k_{a_i}^2 + m_{a_i}^2)^{-1}$, we also summed up with respect to different choices of $a_1<a_2<\ldots <a_n$.  Remaining integral in the formula \ref{rr}  is an expression for a one loop diagram $D^{(n)}_n \left(  m_{a_1}^2, \ldots m_{a_n}^2 ; p_{a_2 a_1}, p_{a_3 a_2}\ldots p_{a_1 a_n}\right)$  for which the number of internal lines is equal to the dimension of space-time. One can say that the reduction formula \ref{rr} is similar to Cutkosky rule \cite{cut}. } On the other hand we want to note that the equality \ref{rr} cannot be integrated, because 
 the momenta, which does not satisfy  conditions {\it A} can contribute. In other words the right hand side of \ref{rr} can have $\delta$-like contributions (of external momenta).
In order to finalize the formulation of reduction formula let us present an explicit form of 'tree' factors of formula \ref{rr}: 
 \begin{eqnarray}   \label{tree}
&  \left[ (k_l^\pm)^2+m_l^2 \right]^{-1} =  \\ \nonumber
& = \det L^{\{a_i\}}  \left[  H_l^ {a_i}  \pm  i \left(  p_{la_1},   p_{a_2 a_1} \otimes p_{a_3 a_1} \otimes \ldots  \otimes  p_{a_n a_1}\right)\sqrt{C^{\{ a_i\}}}  \right]^{-1}
\end{eqnarray}\
 Here the matrix $L^{\{a_i\}}_{a_k a_l} $ has dimension $(n-1)$; both index $a_k$ and $a_l$ run through $a_2, a_3, \ldots a_n$. 
 \begin{eqnarray}   \label{tech}
&  L^{\{a_i\}}_{a_k a_l} = \left(   p_{a_k a_1},  p_{a_l a_1}  \right) \quad ; \\ \label{L}
&  H_l^ {a_i}  = \det L^{\{a_i\}}  \left[  P^2_{la_1} -  \left(  L^{\{a_i\}} \right)^{-1}_{a_i a_k} P^2_{a_k a_1} (p_{la_1},p_{a_ka_1}) \right] , \\ \label{H}
& C^ {\{a_i\}}  =4m^2_{a_1}  \det L^{\{a_i\}} +  P^2_{a_i a_1}  \left(  L^{\{a_i\}} \right)^{-1}_{a_i a_k}  P^2_{a_k a_1}  \det L^{\{a_i\}} , \\ \label{C}
&  P^2_{la_1} = p^2_{la_1}+ m^2_l -m^2_{a_1} .
\end{eqnarray}\
 The vector  $$( p_{a_2 a_1} \otimes p_{a_3 a_1} \otimes \ldots  \otimes  p_{a_n a_1}  )^\mu =\epsilon^{\mu \mu_2 \ldots \mu_n}   p_{a_2 a_1}^{\mu_2} \ldots    p_{a_n a_1}^{\mu_n} $$ is a 'vector product ' of
 $n-1$ vectors in $n$ dimensional space. The $$\left(  p_{la_1},   p_{a_2 a_1} \otimes p_{a_3 a_1} \otimes \ldots  \otimes  p_{a_n a_1}\right)=  \epsilon^{\mu_1 \mu_2 \ldots \mu_n} p^{\mu_1}_{l a_1}  p_{a_2 a_1}^{\mu_2} \ldots    p_{a_n a_1}^{\mu_n} $$  is a volume of parallelepiped spanned by  $n$ vectors $  p_{l a_1},   p_{a_2 a_1}, \ldots  ,  p_{a_n a_1} $.  If conditions {\it A} is valid the determinant 
 \begin{equation} \label{ndg}
 \det L^{\{a_i\}}   = \left( p_{a_2 a_1} \otimes p_{a_3 a_2} \otimes \ldots  \otimes  p_{a_n a_1} \right)^2 \neq 0
  \end{equation}
  does not vanish. These formulae \ref{tree} -\ref{ndg} are derived in section 3, where we also prove that the expression in curly brackets of formula \ref{rr} is a rational function of external momenta and masses.
  We shall use the formulae \ref{tree} -\ref{ndg} in section 2.
  
\section{2. Derivation}
  
  The idea of evaluation of one-loop diagram \ref{Fi}  is to use Cauchy's   residual formula for each component of vector $k$.
  The proof of formula \ref{rr} will go by induction in $n$. The base of the induction is case $n=1$, in this case formula \ref{rr} follows immediately from the standard Cauchy's  formula for evaluation of an integral of a  holomorphic function by residuals. The step of induction will go like this: we shall assume that the formula  \ref{rr} is valid for some $n$ and prove that it is valid also in the dimension $n+1$.
  Let us consider a diagram for $N\geq n+1$:
  \begin{eqnarray}   \label{bi}
& D^{(n+1)}_N \left(  m_1^2, \ldots m_N^2  ;  p_1, \ldots p_N\right)  = \int dk^{n+1}_a \int d^n{\bf k}_a \Pi_{j=1}^N\left( k_j^2+m_j^2 \right)^{-1} =  \nonumber \\
&   =    \int dk^{n+1}_a \int d^n{\bf k}_a \Pi_{j=1}^N\left( {\bf k}_j^2+M_j^2 \right)^{-1} =  \nonumber  \\ 
&  =   \int dk^{n+1}_a D^{(n)}_N \left(  M_1^2, \ldots M_N^2  ;  {\bf p}_1, \ldots  {\bf p}_N\right) 
\end{eqnarray}\
  Here $k=(k^1, \ldots k^n , k^{n+1})  \equiv \left( {\bf k}, k^{n+1}  \right)$ and $M^2_j=m^2_j + (k_j^{n+1})^2 \geq m_j^2$.
  If conditions {\it A} for vectors $p_i$ are valid in $n+1$ dimensional space-time then we can choose the direction of $(n+1)$st axis for vectors $\bf p_i$ to satisfy  conditions {\it A} in $n$ dimensional  space-time.
  So we can use reduction formula \ref{rr} for $D_N^{(n)}$ in the right hand side of formula \ref{bi}.
  \begin{eqnarray}   \label{rri}
&D^{(n+1)}_N = \sum _{\{ a_1<a_2<\ldots <a_n\} }  \left[ \frac{1}{2}\int  d^n {\bf k}_{a_1} \prod_{i=1}^n  \left( {\bf k}_{a_i}^2+M_{a_i}^2 \right)^{-1} \right]  \\ \nonumber
& \times \{ \prod_{l\neq a_i} \left( ({\bf k}_l^+)^2+M_l^2 \right)^{-1} + \prod_{l\neq a_i} \left( ({\bf k}_l^-)^2+M_l^2 \right)^{-1}  \}  
\end{eqnarray}\
  Note that 
  \begin{eqnarray}   \label{rem}
&\int  d^n {\bf k}_{a_1} \prod_{i=1}^n  \left( {\bf k}_{a_i}^2+M_{a_i}^2 \right)^{-1} =   \\ \nonumber
& =D^{(n)}_n \left(  M_{a_1}^2, \ldots M_{a_n}^2  ;  {\bf p}_{a_2 a_1}, \ldots  {\bf p}_{a_1 a_n}\right) \equiv D^{\{ a_i\}}\left(k_{a_1}^{n+1} \right)
\end{eqnarray}\
  The function $D^{\{ a_i\}}\left(k_{a_1}^{n+1} \right)$ depends on $k_{a_1}^{n+1}= k_{a_i}^{n+1}-p^{n+1}_{a_ia_1}$  only by means of $M^2_{a_i}$.
  One can prove that $ D^{\{ a_i\}}$ as function of $k_{a_1}^{n+1}$ has no singularities on the integration contour. We shall represent $ D^{\{ a_i\}}$ in the following way:
  \begin{eqnarray}   \label{rep}
& D^{\{ a_i\}} \left(k^{(n+1)} \right)= \sum _{p=1}^2 D_p^{\{ a_i\}} \left(k^{(n+1)} \right)   \label{sum} \\
&  D_1^{\{ a_i\}} \left(k \right) = \frac{1}{2\pi i} \int_{-\infty}^{\infty} dx \frac{ D^{\{ a_i\}} \left( x \right)}{x-k}, \qquad \Im k >0 \label{Cau}  \\  
&  D_2^{\{ a_i\}} \left(k \right) =  \frac{1}{2\pi i}   \int_{-\infty}^{\infty} dx \frac{ D^{\{ a_i\}} \left( x \right)}{x-k}, \qquad \Im k < 0     \nonumber
\end{eqnarray}\
 For the derivation it will be important that each of functions $ D_p^{\{ a_i\}}  $ depends only  on  $p_{a_2 a_1}, \ldots   p_{a_1 a_n} $ (they depend on $p^{n+1}_{a_{i+1}a_i}$   only by means of  $M^2_{a_i}$)
  ,  on $m_{a_1}, \ldots m_{a_n}$ and on $k^{n+1}_{a_1}$.  The function $ D_1^{\{ a_i\}}  $ is analytic in upper half plane of $k$  and the  function $ D_2^{\{ a_i\}}  $ is analytic in lower half plane .
  Actually one can prove that the functions have joint strip of analyticity: $|\Im k| <\min (m_i)$. In the region of its analyticity each function decay as $k^{-1}$.
  Now let us consider dependence on $k_{a_1}^{n+1}$ of the expression in curly brackets of the formula \ref{rri} :
  \begin{equation} \label{not}
  \{ \prod_{l\neq a_i} \left( ({\bf k}_l^+)^2+M_l^2 \right)^{-1} + \prod_{l\neq a_i} \left( ({\bf k}_l^-)^2+M_l^2 \right)^{-1}  \}   = T_+^{\{ a_i\}} +  T_-^{\{ a_i\}} 
  \end{equation}
  We can see from formulae \ref{tech} -\ref{ndg} that 
   \begin{eqnarray}   \label{com}
&  \left( ({\bf k}_l^+)^2+M_l^2 \right)^{-1}  =  \det L^{\{a_i\}}  \left[  H_l^ {a_i}  \pm  i \left(  p_{la_1},   p_{a_2 a_1} \otimes p_{a_3 a_1} \otimes \ldots  \otimes  p_{a_n a_1}\right)\sqrt{C^{\{ a_i\}}}  \right]^{-1}      \label{expr} \\
&  H_l^ {a_i}  = \det L^{\{a_i\}}  \left[  P^2_{la_1} -  \left(  L^{\{a_i\}} \right)^{-1}_{a_i a_k} P^2_{a_k a_1} (p_{la_1},p_{a_ka_1}) \right] , \qquad P^2_{la_1} = p^2_{la_1}- M^2_l -M^2_{1} \label{ag}  \\  
&   C^ {\{a_i\}}  =4M^2_{a_1}  \det L^{\{a_i\}} +  P^2_{a_i a_1}  \left(  L^{\{a_i\}} \right)^{-1}_{a_i a_k}  P^2_{a_k a_1}  \det L^{\{a_i\}}    \nonumber
\end{eqnarray}\
 All these values depend on $k^{n+1}_{a_1}$ only by means of masses , so: 
 \begin{itemize}
  \item $P^2_{l_{a_1}}$ and $H_l^{\{ a_i\}}$ are linear functions of  $k^{n+1}_{a_1}$
  \item $C^{\{ a_i\}}$ are quadratic  functions of  $k^{n+1}_{a_1}$
   \end{itemize}
  An expression in curly brackets of the formula \ref{not} is an even function of $\sqrt{C^{\{ a_i\}}}$, so it is a meromorphic function $k^{n+1}_{a_1}$.
  Let us find out where the poles of this function. An equation for vanishing of the expression in square  brackets of equation \ref{expr} is quadratic equation with real coefficients, also both signs give the same equation. 
  Due to  conditions {\it A}  we have $\left(  p_{la_1},   p_{a_2 a_1} \otimes p_{a_3 a_1} \otimes \ldots  \otimes  p_{a_n a_1}\right) \neq 0 $. In the next section ( see formula \ref{ineq} ) we prove that $C^{\{ a_i\}}\geq 0$. 
  So the quadratic equation  has two different roots: one in upper half of complex plane another in lower. One is a pole of  the propagator $\left( ({\bf k}_l^+)^2+M_l^2 \right)^{-1}$ and another  
  s a pole of  the propagator $\left( ({\bf k}_l^-)^2+M_l^2 \right)^{-1}$.  In the end of  next section we prove that all the poles are of the first order [$(n+2)$ propagators cannot have a joint pole] .
  Vanishing of the expression in square  brackets of equation \ref{expr}  leads exactly to the joint pole of $(k^2_{a_i}+m^2_{a_i})^{-1}$ here $i=1\ldots n$ and  $(k^2_{l}+m^2_{l})^{-1}$.
  So we can rewrite the integral in formula \ref{rri}  as:
  \begin{eqnarray}   \label{rew}
&D^{(n+1)}_N = \sum _{\{ a_1<a_2<\ldots <a_n\} }     \int  d k^{n+1}_{a_1} D_1^{\{ a_i \}} (k^{n+1}_{a_1} ) \left[  T_+^{\{ a_i\}} (k^{n+1}_{a_1} ) +  T_-^{\{ a_i\}}  (k^{n+1}_{a_1} ) \right]\ +\\ \nonumber
& +   \sum _{\{ a_1<a_2<\ldots <a_n\} }     \int  d k^{n+1}_{a_1} D_2^{\{ a_i \}} (k^{n+1}_{a_1} ) \left[  T_+^{\{ a_i\}} (k^{n+1}_{a_1} ) +  T_-^{\{ a_i\}}  (k^{n+1}_{a_1} ) \right]\ 
\end{eqnarray}\
Let us close the integration contour  in the first term  in upper part of complex plane and in the second in lower. Each of the integrals can be calculated by residuals.  So the value $D^{(n+1)}_N$ can be evaluated (by residuals) by deforming the integration contour to $2(N-n)$  small contours $C^{\pm}$. Each $C^{\pm}$ is a small circle around one of the poles of the integrand. In vicinity of such a pole all other $N-n-1$ propagators can be replaced by their value at the pole and  moved out of the integral .  Now we can deform the integration contour back to real axis. The $D^{(n+1)}_N$ from \ref{rew}   will be represented by a sum of 
  2$N\choose{n+1}$ terms; each terms is a product of $N-n-1$ propagators [which were moved out of the integral] by an integral. The integrand of this integral consists of $n$ propagators $(k^2_{a_i}+m^2_{a_i})^{-1}$
  ($i=1\ldots n$) and a singular factor of  $(k^2_{a_{n+1}}+m^2_{a_{n+1}})^{-1}$ : 
  \begin{eqnarray}   \label{some}
&D^{(n+1)}_N =\sum_{p=1}^2 \sum_{a=\pm}\sum_{\{ a_1<a_2<\ldots <a_n\} }  \prod_{l\neq {a_i}}  (k^2_{l}+m^2_{l})^{-1}  \int  d k^{n+1}_{a_1} D_p^{\{a_1, \ldots a_n \}} (k^{n+1}_{a_1} ) \times \\ \nonumber
& \times \left[ \left( k^a_{a_{n+1}} \right)^2+M^2_{a_{n+1}}\right]^{-1}
\end{eqnarray}\
  It is important that the propagators, which were moved out of the integral are evaluated at a  joint pole of propagators $a_1 \ldots a_{n+1}$:  $$(k^2_{a_i}+m^2_{a_i})=0 \qquad i=1,2, \ldots , n+1 $$
  There are only two solutions $k^\pm_{a_i}$. Now we have to collect all terms at the same tree-like factors. In such a way we arrive at an expression for $D^{(n+1)}_N$:
   \begin{eqnarray}   \label{repr}
&D^{(n+1)}_N \left( m^2_1 , \ldots  , m_N^2 ; p_1 , \ldots , p_N \right) = \\ \nonumber
& \sum_{\{ a_1<a_2<\ldots <a_{n+1}\} } \Phi_1^{\{a_i \}}  \left( m^2_{a_1} , \ldots  , m_{a_{n+1}}^2 ; p_{a_1 a_2} , \ldots , p_{a_{n+1}a_1} \right)  \prod_{l\neq {a_i}} \left[ (\left( k_l^+\right)^2+m^2_{l}\right]^{-1} +  \\ \nonumber
 & + \Phi_2^{\{a_i \}}  \left( m^2_{a_1} , \ldots  , m_{a_{n+1}}^2 ; p_{a_1 a_2} , \ldots , p_{a_{n+1}a_1} \right)  \prod_{l\neq {a_i}} \left[ (\left( k_l^-\right)^2+m^2_{l}\right]^{-1}
\end{eqnarray}\
  By construction an expression for $\Phi_p$ does not depend on $N$. Specifying $N=n+1$ we obtain
 \begin{equation} \label{id}
 \sum_{p=1}^2 \Phi_p^{\{a_i \}}  \left( m^2_{a_1} , \ldots  , m_{a_{n+1}}^2 ; p_{a_1 a_2} , \ldots , p_{a_{n+1}a_1} \right) =\int d^{n+1} k_{a_1} \prod_{i=1}^n ( k^2_{a_i}+m^2_{a_i})^{-1}
   \end{equation}
   just a diagram with $n+1$ internal lines in  $n+1$ space-time.  Now we can use the formula \ref{repr} to obtain:
    \begin{eqnarray}   \label{party}
&D^{(n+1)}_N  -  \sum _{a_1<a_2<\ldots <a_n}  \left[ \frac{1}{2}\int  d^n k_{a_1} \prod_{i=1}^n  \left( k_{a_i}^2+m_{a_i}^2 \right)^{-1} \right] \times  \\ \nonumber
& \times \{ \prod_{l\neq a_i} \left( (k_l^+)^2+m_l^2 \right)^{-1} + \prod_{l\neq a_i} \left( (k_l^-)^2+m_l^2 \right)^{-1}  \}=  \\ \nonumber
& \sum_{\{ a_1<a_2<\ldots <a_{n+1}\} } \Delta^{\{a_i \}}  \left( m^2_{a_1} , \ldots  , m_{a_{n+1}}^2 ; p_{a_1 a_2} , \ldots , p_{a_{n+1}a_1} \right)\times \\ \nonumber
 &\times  \{ \prod_{l\neq {a_i}} \left[ (\left( k_l^+\right)^2+m^2_{l}\right]^{-1} -  \prod_{l\neq {a_i}} \left[ (\left( k_l^-\right)^2+m^2_{l}\right]^{-1} \}
 \end{eqnarray}\
   Here $  \Delta^{\{a_i \}} = \left( \Phi_1^{\{a_i \}}  -\Phi_2^{\{a_i \}} \right)/2$. Note  that $   \Delta^{\{a_i \}}$ depends only on $n$ momenta, since
   $p_{a_1 a_2} + \ldots + p_{a_{n+1}a_1} =0$.
  
   Let us prove now that the right hand side of the formula \ref{party} vanish. We see from formula \ref{tree} for $\left[ (k^\pm_l)^2 +m_l^2 \right]^{-1} $ that curly brackets in 
   the left hand side of formula \ref{party} is a scalar, but the  curly brackets in the right hand side is pseudo-scalar (change the sign at reflection of any axis of momentum space ).
   The whole left hand side of formula \ref{party} is a scalar in $n+1$ dimensional space of momenta. So $\Delta $ in the right hand side have to  be a pseudo-scalar. But it depends only on $n$ linear independent vectors
   $p_{a_i, a_k}$  in $n+1$ dimensional space. This is not possible for pseudo-scalar in $n+1$ dimensional space to depend only  on $n$ linear independent vectors (it should depend on  $n+1$ linear independent vectors). This proves that the right hand side of the formula 
   \ref{party}  is zero. {\it This accomplish the induction}. {\bf We proved the reduction formula \ref{rr}.}
   
   \section{3. Description of the joint pole}
  In this section we study the solution of system of equations \ref{jp} and derive formulae \ref{tree} - \ref{ndg}. First we subtract equation with $i=1$ from other equations in formula \ref{jp}. 
  In such a way we can rewrite system \ref{jp} in the form:
    \begin{eqnarray}   \label{jpl}
&k^2_{a_{1}}+m^2_{a_{1}}=0 \\ \label{kva}
& 2(p_{a_i a_1} ,k_{a_i}) +P^2_{a_i a_1}=0, \qquad i=2,\ldots , n \label{lin}
 \end{eqnarray}\
  This is a system of $n$ equation for $n$ components of the vector $k_{a_{1}}$. Let us look for the solution in the form:
   \begin{equation} \label{der}
k_{a_{1}}^\mu=\sum_{i=2}^n A^{a_1}_{a_i} p_{a_i a_1}^\mu + B^{\{a_{i} \}}  \left( p_{a_2 a_1} \otimes p_{a_3 a_1} \otimes \ldots  \otimes  p_{a_n a_1} \right)^\mu
\end{equation}
  The variable $B^{\{a_{i} \}}$  does  not appear in equations \ref{lin}, so we can consider equations \ref{lin} as a system of $n-1$ equations for defining the value $ A^{a_1}_{a_i}$ ,  $ i=2,\ldots , n$.
 If conditions   {\it A}  for external momenta   are valid  then the determinant of the system \ref{lin} does not vanish, see formula  \ref{ndg}. So the system has unique solution see formulae \ref{tech}- \ref{C}:
 $$   A^{a_1}_{a_i} =-\frac{1}{2} \left(  L^{\{a_i\}} \right)^{-1}_{a_i a_k} P^2_{a_1 a_k} $$
 Now we can use formula \ref{jpl} to find 
 $$  B^{\{a_{i} \}} =\pm \frac{i}{2\det L^{\{ a_i\}}}\sqrt{C^{\{ a_i\}}} $$
 Now we can write the solution of the system \ref{jp} :
   \begin{eqnarray}   \label{sol}
&\left(k^\pm_{a_1} \right)^\mu = -\frac{1}{2} \left(  L^{\{a_i\}} \right)^{-1}_{a_i a_k} P^2_{a_k a_1} p^\mu_{a_i a_1} \pm \\ \nonumber
& \pm \frac{i}{2\det L^{\{ a_i\}}}\sqrt{C^{\{ a_i\}}}  \left( p_{a_2 a_1} \otimes \ldots  \otimes p_{a_{n}a_1} \right)^{\mu} \\ \nonumber
 \end{eqnarray}\
 The formula \ref{tree} for tree factors  follows from here. The value $C^{\{ a_i\}}$ does not depend on $l$, it is a polynomial of the external momenta.
 Also both of tree factors in curly brackets of the formula \ref{rr} contain $\sqrt{C^{\{ a_i\}}}$ , the sum is even function of  $\sqrt{C^{\{ a_i\}}}$ and does not contain square root singularity. 
 So under conditions {\cal A} the system \ref{jp} has two solutions \ref{sol} , they are different  $k^+_{a_1} \neq {k^-_{a_1}}$  and complex conjugated $k^+_{a_1}= \overline{k^-_{a_1}}$.
 We conclude this because the matrix $ L^{\{a_i\}} $ is positive,  so 
\begin{equation} \label{ineq}
C^{\{ a_i\}} \geq 4m^2_{a_1} \det  L^{\{a_i\}}
 \end{equation}
 and $ B^{\{a_{i} \}} \neq 0$.  \par
{ \bf  Now let us prove that under conditions {\cal A} any of   $(n+1)$ propagators in $n$ dimensional space-time cannot have joint pole. }
 Let us assume that on top of equations \ref{jp} one more equation is valid 
 $$k^2_b + m^2_b=0= \left(k_{a_1} +p_{ba_1}  \right)^2  +  m^2_b, \quad \mbox{here}  \quad b\neq a_i  \quad \mbox{for} \quad  i=1, \ldots , n$$
 Let us subtract from here the equation \ref{jpl} : we shall obtain a system of $n$ linear equations for $A^{a_1}_{a_i}$ and $ B^{\{a_{i} \}}$:
 $$ 2(p_{a_i a_1} ,k_{a_i}) +P^2_{a_i a_1}=0 , \qquad  2(p_{b a_1} ,k_{a_i}) +P^2_{a_1 b}=0  $$
 This system has unique solution and $ B^{\{a_{i} \}}$ is real, but the initial system \ref{jpl} -\ref{lin} gave  pure imaginary $ B^{\{a_{i} \}}$. This contradiction proves the statement. {\it In such a way we proved that all the  poles in formula 
 \ref{Fi} are simple.} \par
 
 The authors are grateful to L.D. Faddeev for discussions.

\end{document}